\documentclass[fleqn,twoside]{article}
\usepackage{espcrc2}
\usepackage{epsfig}

\usepackage{graphicx}
\title{Pion scattering on the lattice with chirally improved
fermions\thanks{Poster presented at LATTICE 2004. 
This work was supported by DFG and BMBF.}}

\author{{Christof Gattringer, Dieter Hierl, and
Rainer Pullirsch\\
(for the BGR [Bern-Graz-Regensburg] Collaboration)} 
\vskip5mm
Institut f{\"u}r Theoretische Physik, Universit{\"a}t
Regensburg, 93040 Regensburg, Germany.
\vskip3mm}
       
\begin{document}
\begin{abstract}
We report preliminary results for the $I=2$ pion scattering length
$a_0$, calculated with chirally improved fermions. 
After chiral extrapolation our results are in good agreement
with both, experimental and theoretical predictions.
\vspace{1pc}
\end{abstract}
\maketitle
\section{Introduction}
Since the pion is the goldstone boson of the spontaneous breaking of 
chiral symmetry, observables related to the pion give insight into the 
chiral dynamics of QCD. For example the pion scattering length is 
a parameter entering the effective chiral lagrangian
of QCD and a precise determination of this observable in an ab-initio 
lattice calculation is desirable. 
On the lattice scattering data of stable particles
can be calculated using L\"uscher's formula \cite{Lus1} which 
relates the scattering length to the energy $W$ of the scattering state. 
The crucial technical challenge is a very precise determination of 
$W$ and advanced strategies for error reduction have to be used.  
For chirally sensitive quantities, such as the pion
scattering length, it is important to be able to work with small 
pion masses. Using fermions based on the Ginsparg-Wilson equation 
it is now possible to reach pion masses much lower than what was 
accessible with more traditional formulations. This implies that 
the chiral extrapolation of the results to the physical pion mass 
becomes more reliable. 

\section{Setting of our calculation}
Our calculation is based on the chirally improved (CI) Dirac operator
\cite{Gat1}. CI fermions are a sys\-te\-ma\-tic approximation of a solution
of the Ginsparg-Wilson equation and have been tested 
successfully \cite{BGR1} in
spectroscopy down to pion masses of ${m_\pi=265}$~MeV.

\begin{table}[t]
\begin{center}
\hspace*{-5.25pt}
\begin{tabular}{cccccc}
\hline
ensemble &$\!\!\!\beta$ & $\!\!\!a$(fm) & $\!\!\!L$(fm) & size & conf. \\\hline
A & $\!\!\!7.90$ & $\!\!\!0.15$ & $\!\!\!2.4$ & $16^3 \! \times \! 32$ & 97 \\
B & $\!\!\!7.90$ & $\!\!\!0.15$ & $\!\!\!1.8$ & $12^3 \! \times \! 24$ & 100 \\
C & $\!\!\!8.35$ & $\!\!\!0.10$ & $\!\!\!1.6$ & $16^3 \! \times \! 32$ & 91 \\\hline
\end{tabular}
\end{center}
Table 1. Parameters and statistics of our quench\-ed ensembles.
\vspace{-5mm}
\end{table}
We use gauge configurations generated with the L\"uscher-Weisz action.
Our quark masses are in the range of $am = 0.015$ to $am = 0.2$.  
The scale was determined in \cite{scale} using
the Sommer parameter. The parameters of the three ensembles we used are listed 
in Table 1.

\section{L\"uscher's formula}
If two stable particles are confined in a box of size $L$, the
energy $W$ of the corresponding scattering state can be expanded 
\cite{Lus1} in a power series in $L^{-1}$. The
coefficients of this series are related to the elastic scattering
amplitude,
\begin{equation}
W - 2m_\pi = - \frac{4\pi\,a_0}{m_\pi\,L^3}\!\left[1+
c_1\frac{a_0}{L}+c_2\left(\frac{a_0}{L}\right)^2\right] 
... \, , \; 
\label{lusch}
\end{equation}
where the omitted terms are ${\cal O}(L^{-6})$ and $c_1$, $c_2$ 
are numerical constants that are known analytically
($c_1=-2.837297$, $c_2=6.375183$).
Thus, for calculating the s-wave scattering length $a_0$, one has
to compute $m_\pi$ and $W$ on the lattice and to solve the cubic 
equation (\ref{lusch}) for $a_0$. The challenge is to obtain the
difference $W - 2m_\pi$ with high accuracy since the solution
of the cubic equation is very sensitive to this difference.

\section{Two pion scattering state energy}
We calculate the pion mass $m_\pi$ from a 2-parameter fit of the pseudoscalar
correlator ($T$ denotes the temporal extent of the lattice)
\begin{equation}
\langle C_2(t) \rangle = \langle\widehat{P}(t)P^\dagger(0)\rangle
\sim A \cosh(m_\pi[T/2-t]),
\label{g2}
\end{equation} 
where $P^\dagger$ generates $I=1$ pseudoscalar mesons and $\widehat{P}$
denotes the zero-momentum Fourier transform of the corresponding 
annihilator. 

The energy $W$ of the scattering states can be extracted from 
\begin{equation}
\langle C_4(t) \rangle \;=\;\langle\widehat{P}(t)\widehat{P}(t)P^\dagger(0)
P^\dagger(0)\rangle \; . 
\label{g4}
\end{equation}
In both, the 2- and the 4-point correlators, we use Jacobi
smeared quark sources and sinks. We also experimented with wall-type sources
but found no improvement of the signal. 

While the pion mass can be extracted quite accurately from (\ref{g2}), a 
direct determination of the scattering state 
energy $W$ from a fit to (\ref{g4})
is unsatisfactory. A successful method \cite{Jug1} to reduce the statistical
noise is to consider the ratio 
\begin{equation}
\left\langle \frac{C_4(t)}{C_2(t)^2} \right\rangle \sim 
\frac{A + B \cosh(W[T/2-t])}{\cosh(m_\pi[T/2-t])^2} \; ,
\label{ratio}
\end{equation}
where fluctuations of the correlators on single configurations cancel. 
We first determine $m_\pi$ from (\ref{g2}) and subsequently use 
this value in the 3-parameter fit (\ref{ratio}) which determines $W$.
Our fits are fully correlated and errors were determined with the jackknife 
method.

\section{Chiral extrapolation} 
For the chiral extrapolation of our data we use chiral perturbation theory.
In one-loop chiral perturbation theory one finds \cite{Gas1}
\begin{eqnarray}
a_0\!\!&\!\!=\!\!&\!\!-\frac{m_\pi}{16\pi\,f_0^2}\!
\left[1-\frac{m_\pi^2}{f_0^2}\!\left(G_\Lambda-\frac{7
\ln(m_\pi^2)}{32\pi^2}+\dots\right)\!\right].\nonumber\\
\label{a0func}
\end{eqnarray}
$G_\Lambda$ is a (scale dependent) combination of LEC's and 
$f_0$ is the pion decay constant in the chiral limit. However, we do not
fit our data directly to the form (\ref{a0func}), but consider two 
combinations of $a_0$ with $m_\pi$.
To remove the leading $m_\pi$-dependence of the scattering length 
we study the ratio
\begin{equation}
F_1(m_\pi^2)\; = \;\frac{a_0}{m_\pi}\;.
\end{equation}
\begin{figure}[t]
\centerline{\epsfig{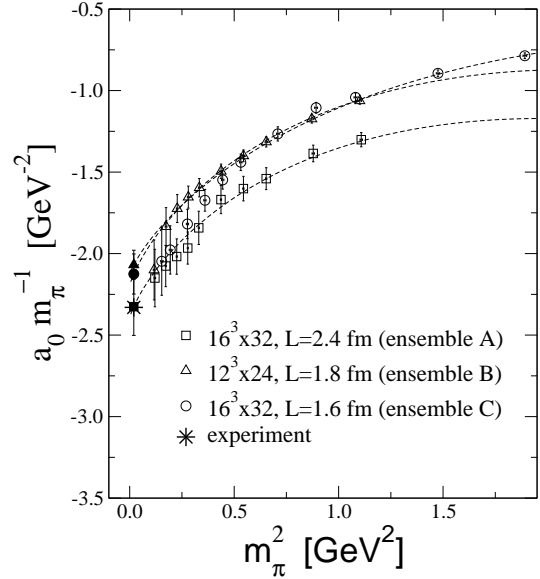}}
\vspace{-8mm}
\caption{Results for $F_1(m_\pi^2) = a_0/m_\pi$ as a function of the 
squared pion mass. The open symbols are our data, the filled symbols 
their extrapolation to the physical pion mass. The curves represent 
our fits. The experimental value is indicated by the asterisk. 
\label{f1plot}}
\end{figure}
We also consider the product 
\begin{equation}
F_2(m_\pi^2)\;=\;a_0\,m_\pi \; ,
\end{equation}
which is often used by experiments. We perform the chiral 
extrapolation for both $F_1$ and $F_2$ using the accordingly modified 
right hand side of (\ref{a0func}). 
We furthermore experimented with adding higher order terms from chiral 
perturbation theory but did not observe a large effect.

\section{Results}
In Fig.\ 1 we present our results for the ratio $F_1$. The open symbols 
are our data and the filled symbols are their extrapolation to the physical 
pion mass. The asterisk represents the experimental value. When comparing the 
two ensembles B (triangles) and C (circles) 
which are similar in size ($L = 1.8$ fm and $L = 1.6$ fm) but have a different 
lattice spacing ($a = 0.15$ fm and $a = 0.1$ fm) we find, at least for larger
pion masses, good agreement of the data points.
This indicates that discretization errors are small, similar to 
what was already found in quenched hadron spectroscopy with the chirally 
improved Dirac operator \cite{BGR1}. A comparison with the data obtained on our 
larger volume (ensemble A, $L = 2.4$ fm, $a = 0.15$ fm) shows, however,
that at the smaller volume with $L = 1.8$ fm we still observe a finite size effect.   

\begin{figure}[t]
\centerline{\epsfig{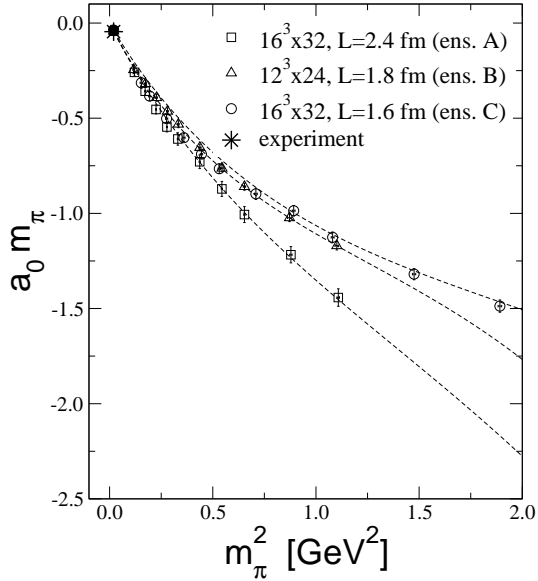}}
\vspace{-8mm}
\caption{Same as Figure 1, but now for the product 
$F_2(m_\pi^2) = a_0 \, m_\pi$.
\label{f2plot}}
\end{figure}

The same picture also holds for the analysis of $F_2$. Again the two 
ensembles with small physical volume but different lattice spacing agree
reasonably well, indicating that scaling violations are small. The larger 
volume gives lower values for $F_2$, showing that at $L=1.8$ fm finite
size effects cannot be neglected.  

Because of the good scaling and the finite size effect we expect the 
results from ensemble A ($16^3 \times 32, a = 0.15$ fm, $L = 2.4\,fm$) 
to be our best estimate. For this ensemble we list in Table 2 our results 
for $a_0 m_\pi$ obtained from the
extrapolation of both $F_1(m_\pi^2)$ and $F_2(m_\pi^2)$ to the physical pion
mass. We compare our final results to recent experimental data \cite{Pis1} and
to a two-loop result from chiral perturbation theory \cite{Col1}. Given the
fact that our calculation is quenched, the agreement of our results with the 
experimental number is surprisingly good. 

\begin{table}[t]
\begin{center}
\hspace*{-11.2pt}
\begin{tabular}{cccc} \hline
extrap. & our data & experiment & $\chi$-PT \\ \hline
$F_1(m_\pi^2)$ & -0.0453(35) & -0.0454(34) & -0.0444(10) \\
$F_2(m_\pi^2)$ & -0.0425(97) & -0.0454(34) & -0.0444(10) \\
\hline
\end{tabular}
\end{center}
Table 2. Our final results for $a_0 m_\pi$ compared to experimental data 
and to 2-loop chiral perturbation theory.
\vspace{-5mm}
\end{table}

\section*{Acknowledgement}
We thank Jimmy Juge for important comments and discussions.
Our computations were done on the Hitachi SR8000 at LRZ in Munich 
and at the RZ Regensburg.

\end{document}